\documentclass[runningheads]{llncs}

\usepackage{booktabs} 
\usepackage{multirow}
\usepackage{changes}
\usepackage{adjustbox}
\usepackage{tablefootnote}
\usepackage[utf8]{inputenc} 
\usepackage{graphicx}
\usepackage{hyperref}
\usepackage{microtype}
\usepackage{enumitem}
\renewcommand{\paragraph}[1]{\noindent {\bf #1}}

\definechangesauthor[color=purple]{MS}

\usepackage{array}

\makeatletter
\newcommand{\thickhline}{%
    \noalign {\ifnum 0=`}\fi \hrule height 1pt
    \futurelet \reserved@a \@xhline
}
\newcolumntype{"}{@{\hskip\tabcolsep\vrule width 1pt\hskip\tabcolsep}}
\makeatother

\begin{document}
\title{Open Research Knowledge Graph:\\A System Walkthrough}


\author{Mohamad Yaser Jaradeh\inst{1,2}\and
Allard Oelen\inst{1,2}\and
Manuel Prinz\inst{2}\and
Markus Stocker\inst{2} \and
S\"oren Auer\inst{2,1}} 

\institute{L3S Research Center, Leibniz University of Hannover, Germany
\email{\{jaradeh,oelen\}@l3s.de} \and 
TIB Leibniz Information Centre for Science and Technology, Germany
\email{\{manuel.prinz,markus.stocker,auer\}@tib.eu}} 

\authorrunning{M.Y. Jaradeh et al.}

\maketitle

\begin{abstract}
Despite improved digital access to scholarly literature in the last decades, the fundamental principles of scholarly communication remain unchanged and continue to be largely document-based. Scholarly knowledge remains locked in representations that are inadequate for machine processing. The Open Research Knowledge Graph (ORKG) is an infrastructure for representing, curating and exploring scholarly knowledge in a machine actionable manner. We demonstrate the core functionality of ORKG for representing research contributions published in scholarly articles. A video of the demonstration \cite{tibav:42537} and the system\footnote{\url{https://orkg.org/orkg/}} are available online.

\end{abstract}


\setlist{nosep}

\section{Introduction}
Documents are central to scholarly communication. Virtually all research findings are nowadays communicated by means of electronic scholarly articles. Scholarly knowledge communicated in such form is hardly accessible to computers and the primary machine-supported tasks are largely limited to traditional full-text search. As such, the current scholarly infrastructure does not exploit modern information systems and technologies to their full potential \cite{journals/isf/Hars01}.

We argue that there is an urgent need for a more flexible, fine-grained, context sensitive representation of scholarly knowledge and thus corresponding infrastructure for knowledge curation, publishing and processing. Furthermore, we suggest that representing scholarly knowledge as structured, interlinked, and semantically rich knowledge graphs is a key element of a technical infrastructure~\cite{auer18towards}.

While some important conceptual foundations have been developed over several decades \cite{Allen2011,journals/isf/Hars01}, knowledge graph infrastructure for science has recently gained momentum in the literature and community. The Research Graph \cite{aryani17researchgraph} is a prominent example of an effort that aims to link publications, datasets, and researchers. The Scholix project \cite{burton17scholix} standardized the information about the links between scholarly literature and data exchanged among (primarily) publishers and data repositories. More recently, the FREYA H2020 project\footnote{\url{https://project-freya.eu}} has released information on their work towards a PID Graph \cite{fenner19pidgraph}. The key distinguishing factor between these systems and the ORKG is the granularity of captured scholarly knowledge (article bibliographic metadata vs. materials, methods, and results communicated in articles).

\section{Architecture and Features}
The ORKG leverages knowledge graph technologies to represent, store, link, and process scholarly knowledge. It has two main components: The back end, which contains the logic to handle requests by client applications and the front end through which users create, curate or explore scholarly knowledge.

The concept of \texttt{ResearchContribution} is central to the ORKG as it represents key aspects of scholarly knowledge in structured, machine actionable form. A \texttt{ResearchContribution} is an information object which relates the \texttt{ResearchProblem} addressed by the contribution with a \texttt{ResearchMethod} and at least one \texttt{ResearchResult}.

\begin{figure}[tb]
    \centering
	\includegraphics[width=0.5\columnwidth]{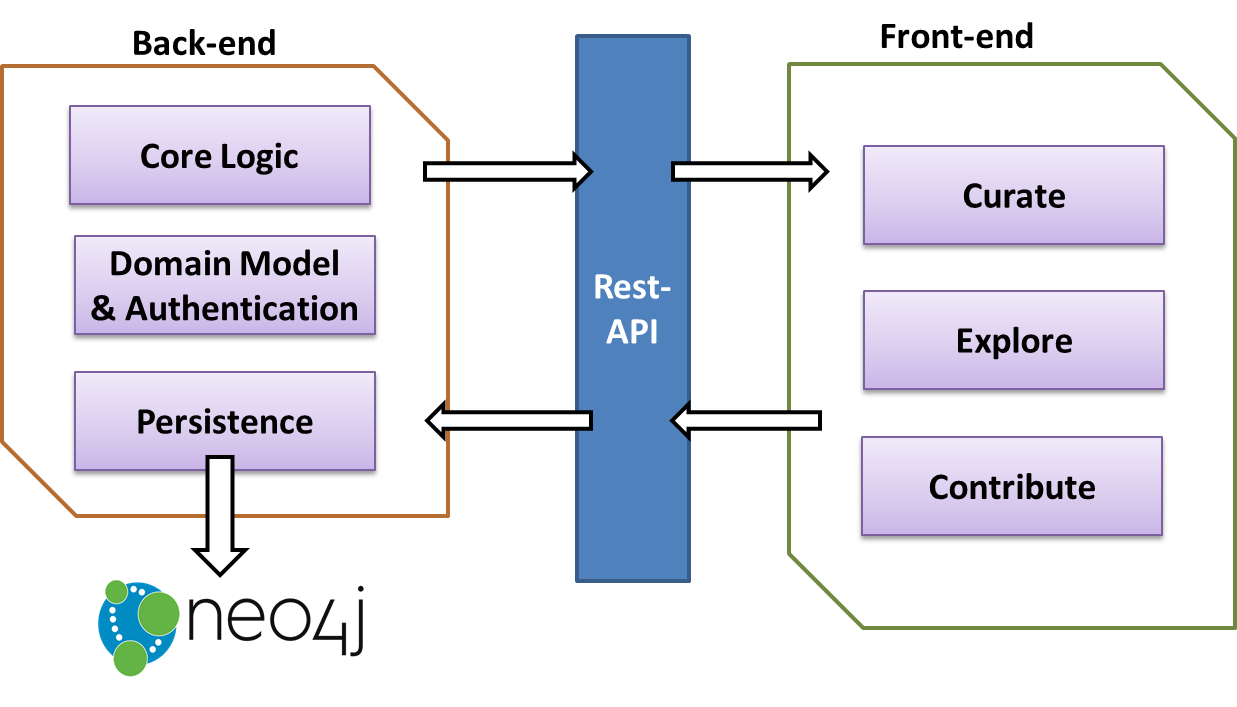}
	\caption{The ORKG architecture showing the main infrastructure components.}
	\label{fig:architecture}
\end{figure}

The ORKG back end represents descriptions by means of a graph data model. Similarly to the Research Description Framework\footnote{\url{https://www.w3.org/RDF/}} (RDF), the data model is centered around the concept of a statement, a triple consisting of two nodes (resources) connected by a directed edge. In contrast to RDF, it allows annotating edges and statements. As metadata of statements, provenance information, e.g. when and by whom a statement was created, is a concrete and relevant application of such annotations.

ORKG users interact with the front end (UI), which guides users through the process of creating research contribution descriptions in a step by step manner. More advanced features of the infrastructure include the ability to directly find similar contributions (and related papers), thus enabling efficient state-of-the-art comparison and literature review. Figure \ref{fig:architecture} depicts the ORKG system architecture.

\begin{figure}[tb]
    \centering
    \includegraphics[width=.9\textwidth]{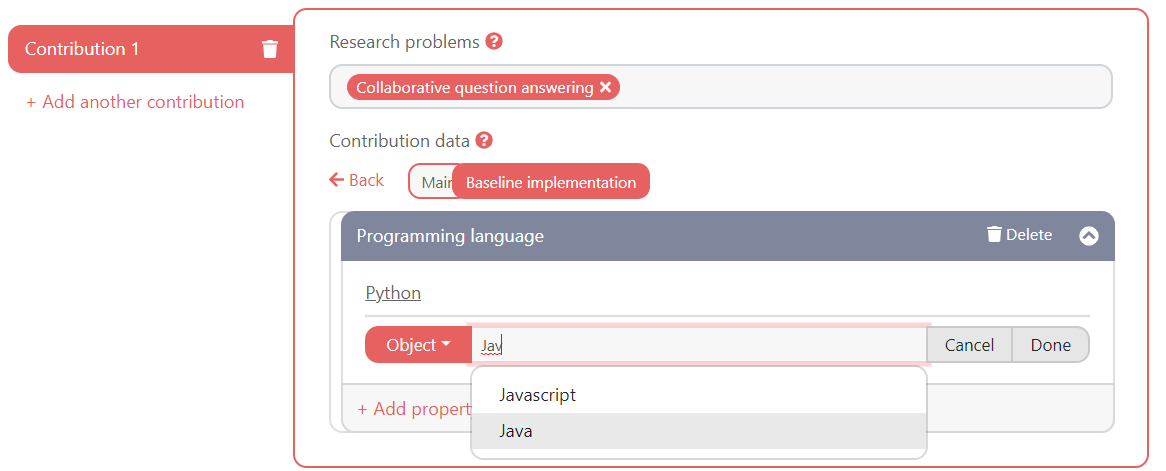}
    \caption{ORKG UI curation wizard step (3) depicting the auto-completion feature that enables linking existing resources (here, Java).}
    \label{fig:demo_1}
\end{figure}

\section{Use Case}
Consider the following research contribution: \textit{FRANKENSTEIN~\cite{Singh:2018:WRW:3178876.3186023} is a collaborative question answering (QA) framework written in Java and Python. It generates QA pipelines based on predictions for the best performing pipelines obtained via a supervised learning model. FRANKENSTEIN evaluates the results against QALD and LC-Quad datasets using the f1-score and accuracy@k metrics}. We can identify the following instances of relevant concepts:

\begin{itemize}
    \item \texttt{Problem}: Collaborative question answering
    \item \texttt{Programming Language}: Python, Java
    \item \texttt{Approach}: Generate optimal QA pipelines
    \item \texttt{Datasets}: QALD, LC-Quad
    \item \texttt{Evaluation Metrics}: f1-score, accuracy@k
\end{itemize}

\noindent
Using the ``Add paper'' wizard (Figure \ref{fig:demo_1}), we can create structured descriptions that encode, in machine actionable manner, the key information of research contributions. This process is straightforward also for non-technical users. Firstly, bibliographic metadata is collected, either via DOI lookup using the Crossref API or manually. Secondly, users can classify their paper according to the research domain. Finally, the research contributions described in the paper are collected using a flexible and dynamic interface. 

\begin{figure}[tb]
    \centering
    \includegraphics[width=.85\textwidth]{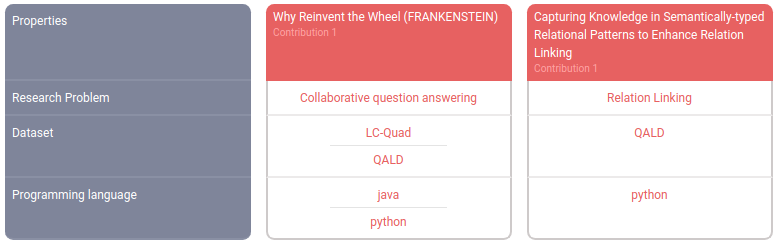}
    \caption{ORKG UI state-of-the-art comparison for research contributions, showing a subset of shared properties between two articles.}
    \label{fig:demo_2}
\end{figure}

\section{Conclusion and Future Work}
We presented the Open Research Knowledge Graph, an infrastructure that makes the first steps of a larger research and development agenda that aims to transition document-based scholarly communication to a knowledge-based information representation.
In future work, we will include additional techniques from machine support to content creation and curation (such as NLP tools to suggest/annotate relevant concepts on behalf of users). Furthermore, we will further develop novel features such as state-of-the-art comparisons (Figure \ref{fig:demo_2}). Such features will underscore the possibilities enabled by machine actionable scholarly knowledge and corresponding infrastructure.

\footnotesize
\section*{Acknowledgment}
This work has received funding from the European Research Council (ERC) under the European Union’s Horizon 2020 Research and Innovation Programme (Grant agreement No. 819536).

\small
\bibliographystyle{splncs04}
\bibliography{bibliography}

\end{document}